\documentclass[showpacs,aps,graphicx,]{revtex4}
\usepackage{amsmath}
\usepackage{mathrsfs}
\usepackage{amsfonts}
\usepackage{amssymb}
\usepackage{graphicx}
\usepackage{caption}
\usepackage{eufrak}
\usepackage{multirow}
\usepackage{float}
\usepackage[colorlinks,linkcolor=blue,anchorcolor=blue,citecolor=blue]{hyperref}
\usepackage{soul,color,xcolor}
\usepackage{graphicx}%
\usepackage{epstopdf}
\usepackage{multirow}%
\soulregister\upcite7
\soulregister\citep7
\soulregister\citet7
\soulregister\ref7
\soulregister\pageref7
\soulregister\and7
\soulregister\Name7

\begin{document}

\title{Bidirectional controlled quantum state preparation in high-dimensional quantum system}

\author{Si-Qi Du, Hai-Rui Wei\footnote{Corresponding author: hrwei@ustb.edu.cn}}

\address{School of Mathematics and Physics, University of Science and Technology Beijing, Beijing 100083, China}

\date{\today }

\begin{abstract}
High-dimensional quantum system exhibits unique advantages over the qubit system in some quantum information processing tasks.
We present a program for implementing deterministic bidirectional controlled remote quantum state preparation (BCRSP) in arbitrary $N$-dimensional (quNit) system.
By introducing two generalized Greenberger-Horne-Zeilinger (GHZ) states as quantum channels, two communication parties can simultaneously prepare a single-particle high-dimensional state at each other's site under the control of Charlie.
Compared with the previous counterparts, the significant advantage of our scheme is that the high-dimensional CNOT operations are not required.
Moreover, the performance our scheme are evaluated. The evaluation of the performance shows that if the quNit is encoded in the spatial mode of single photons, our scheme can be accomplished solely  using only linear optical elements.

keywords: remote quantum state preparation, high-dimensional system, generalized Greenberger-Horne-Zeilinger (GHZ) states, bidirectional controlled remote quantum state preparation

\end{abstract}


\maketitle

\newcommand{\upcite}[1]{\textsuperscript{\textsuperscript{\cite{#1}}}}

\section{Introduction}\label{sec1}

Remote state quantum preparation (RSP), is one of the important branches of quantum communication.
RSP was first proposed by Lo et al.\upcite{lo2000classical} which remotely transmit a known pure single-qubit state from Alice (one site) to Bob (another site) via a prior shared maximally entangled state with the help of some classical feed-forward operations.
In RSP process, transmitted state is only completely known by the sender, and does not owned to the sender.
Later, great progress has been made in RSP both in theoretical and experimental, and various interesting RSPs were proposed using many kinds of methods, such as continuous variable RSP,\upcite{variable1,variable2,variable3} RSP of multipartite pure states,\upcite{multipartitepures1,multipartitepures2,multipartitepures3}
RSP of high-dimensional pure states,\upcite{highpure1,highpure2} RSP of mixed states,\upcite{mixed1,mixed2} cyclic RSP,\upcite{cyclic1} joint RSP,\upcite{joint1,joint2,xia2007multiparty} deterministic RSP,\upcite{deterministic1,deterministic2,deterministic3} controlled RSP,\upcite{controlled1,controlled2,controlled3,controlled4} and bidirectional controlled RSP.\upcite{bidirectional1,chen2017controlled}

The original bidirectional controlled remote state quantum preparation (BCRSP) was proposed by Cao et al.\upcite{cao2013deterministic} in 2013.
In their scheme, two remote parties are allowed to simultaneously transmit unknown quantum states to each other and reconstruct the state received by them under the control of the third party (supervisor Charlie).
Later, many improved BCRSPs were proposed to meet the needs of different communication scenarios,\upcite{chen2017controlled,li2019bidirectional,probabilistic1,ma2020deterministic}
such as probabilistic BCRSP,\upcite{probabilistic1} deterministic BCRSP,\upcite{ma2020deterministic} hybrid BCRSP,\upcite{hybrid1,hybrid2} joint BCRSP,\upcite{JBCRSP1,JBCRSP2} symmetric BCRSP,\upcite{zhang2016deterministic,symmetric} asymmetric BCRSP,\upcite{ma2017asymmetric,sun2019asymmetric,asymmetric} and cyclic controlled RSP.\upcite{cycliccontrol1,cycliccontrol2}
All above schemes are focused on qubit systems, and six-qubit, ten-qubit, or eleven-qubit maximally entangled channels are required.

High-dimensional quantum systems offer many remarkable advantages over two-dimensional systems, including enhanced channel capacity, higher noise resistance, improved efficient quantum simulation and computation, reduced circuit complexity, larger violation of Bell inequality, simplified experimental setup,\upcite{highdimensional1,highdimensional2,highdimensional3,highdimensional4,highdimensional5,highdimensional6,highdimensional7,highdimensional8} etc.
However, high-dimensional BCRSPs are rarely investigated both theoretically and experimentally. In 2019, Ma et al.\upcite{bai2019bidirectional} proposed a BCRSP in qutrit (three-dimensional) system via two generalized Greenberger-Horne-Zeilinger (GHZ) states, and six two-qutrit CNOT gates are required. In 2021, Xu et al.\upcite{Xu2021bidirectional} designed a scheme for implementing a deterministic controlled RSP of two-qutrit equatorial states by using one generalized Bell state and one generalized three-qutrit GHZ state. In this scheme, two two-qutrit CNOT gates are necessary. In 2023, Jiang et al.\upcite{shi2023controlled} proposed a scheme to achieve BCRSP in four-dimensional system via two generalized GHZ states and eight two-qudit CNOT gates. Recently, cyclic controlled RSP in three-dimensional system has been presented.\upcite{shi2021controlled,ma2022cyclic} Above works mainly are focused on a type of equatorial quantum state, which is widely used in quantum key distribution,\upcite{equatorial1} quantum error correction,\upcite{equatorial2} and quantum state transmission.\upcite{equatorial3}

In this paper, we first proposed two schemes to implement BCRSPs in 3D and 4D systems with two generalized GHZ states as quantum channels.
In our schemes, the two remote parties Alice and Bob are both sender and receiver. By introducing two pre-shared generalized GHZ states and feed-forward
measurement strategy, the two parties simultaneously prepare a single-particle three-dimensional (four-dimensional) state at each other's site under the controller's (Charlie) control.
In previous single-particle three-dimensional (four-dimensional) BCRSP counterparts,\upcite{bai2019bidirectional,shi2023controlled} six (eight) high-dimensional CNOT gates are employed. The high-dimensional CNOT gates are not required in our schemes.
Subsequently, we extend the scheme to arbitrary $N$-dimensional (quNit) system via only two generalized three-quNit GHZ states.
Lastly, we evaluate the performances of our schemes. The result indicates that by encoding quNit in the spatial of single photons, our scheme can be accomplished by solely using linear optical elements.

\section{Bidirectional controlled quantum state preparation in high-dimensional system} \label{sec2}

\subsection{Single-particle three-dimensional state}\label{sec21}

\begin{figure}[htbp]
\centering
\includegraphics[width=13 cm]{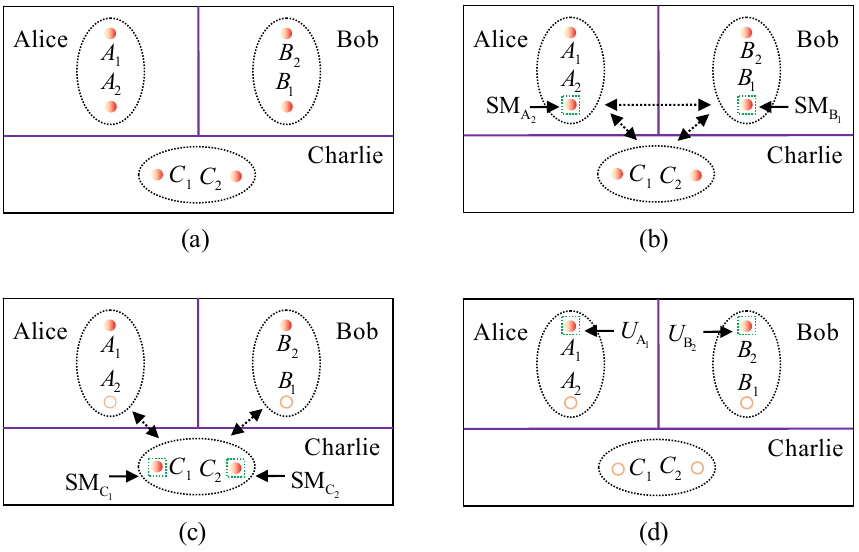}
\caption{
The processes of BCRSP by using two GHZ states as quantum channels. The whole process of the project have four parts. (a) Dashed dots represent particles owned by three communicators. (b) Alice and Bob make single-particle measurements on particles $A_2$ and $B_1$, respectively. Then Alice (Bob) shares her (his) measurement result with Bob (Alice) and Charlie through the quantum channel.
(c) Charlie makes single-particle measurements on his particles $C_1$ and $C_2$. Then he informs his measurement results to Alice and Bob. (d) Alice and Bob reshape the target particles by performing classical feed-forward operations on their particles $A_1$ and $B_2$.}
\label{Fig.1}
\end{figure}

Assuming that Alice wishes to prepare a single qutrit (three-dimensional) state remotely in Bob's site under the control of Charlie. The transmitted state is given by
\begin{eqnarray}                         \label{1}
  \begin{split}
    |\chi\rangle_A = \frac{1}{\sqrt{3}}(|0\rangle + e^{\texttt{i}\delta_1}|1\rangle + e^{\texttt{i}\delta_2} |2\rangle)_A.
  \end{split}
\end{eqnarray}
At the same time, Bob helps Alice prepare another single-qutrit state written as
\begin{eqnarray}                         \label{2}
	\begin{split}
    |\tilde{\chi}\rangle_B = \frac{1}{\sqrt{3}}(|0\rangle + e^{\texttt{i}\tilde{\delta}_1}|1\rangle + e^{\texttt{i}\tilde{\delta}_2}|2\rangle)_B.
  \end{split}
\end{eqnarray}
Here $\delta_1$, $\delta_2$, $\tilde{\delta} _1$ and $\tilde{\delta}_2$ are real.  $\delta_1$ and $\delta_2$ ($\tilde{\delta}_1$ and $\tilde{\delta}_2$) are completely known to Alice (Bob) but unknown to Bob and Charlie (Alice and Charlie).
In order to complete above task, Alice, Bob and Charlie pre-share two three-qutrit generalized GHZ states, which can be expressed as
\begin{eqnarray}                         \label{3}
  \begin{split}
    |\Psi\rangle = \frac{1}{\sqrt{3}}(|000\rangle + |111\rangle + |222\rangle)_{A_1B_1C_1}\otimes\frac{1}{\sqrt{3}}(|000\rangle + |111\rangle + |222\rangle)_{A_2B_2C_2}.
  \end{split}
\end{eqnarray}
We consider Alice owns qutrit pair ($A_1$, $A_2$), Bob holds qutrit pair ($B_1$, $B_2$), and Charlie processes qutrit pair ($C_1$, $C_2$).
The basic processes of BCRSP are shown in the Figure \ref{Fig.1}.

Firstly, Alice carries out single-particle measurement on her particle $A_2$ with mutually orthogonal basis vector
   $\{|\varphi_0\rangle_{A_2}, |\varphi_1\rangle_{A_2}, |\varphi_2\rangle_{A_2}\}$.
Here
\begin{eqnarray}                         \label{4}
\begin{split}
   &|\varphi_0\rangle_{A_2} = \frac{1}{\sqrt{3}}(|0\rangle + e^{-\texttt{i}\delta _1}|1\rangle + e^{-\texttt{i}\delta _2}|2\rangle)_{A_2}, \\
   &|\varphi_1\rangle_{A_2} = \frac{1}{\sqrt{3}}(|0\rangle + e^{\frac{\texttt{i}2\pi}{3}}  e^{-\texttt{i}\delta _1}|1\rangle
                                                           + e^{\frac{\texttt{i}4\pi}{3}}  e^{-\texttt{i}\delta _2}|2\rangle)_{A_2},\\
   &|\varphi_2\rangle_{A_2} = \frac{1}{\sqrt{3}}(|0\rangle + e^{\frac{\texttt{i}4\pi}{3}}  e^{-\texttt{i}\delta _1}|1\rangle
                                                           + e^{\frac{\texttt{i}2\pi}{3}}  e^{-\texttt{i}\delta _2}|2\rangle)_{A_2}.
\end{split}
\end{eqnarray}
At the same time, Bob performs a single-particle measurement on his particle $B_1$ with the basis $\{ |\tilde{\varphi}_0\rangle_{B_1}, |\tilde{\varphi}_1\rangle_{B_1}, |\tilde{\varphi}_2\rangle_{B_1} \}$. Here
\begin{eqnarray}                         \label{5}
  \begin{split}
	   &|\tilde{\varphi}_1\rangle_{B_1} = \frac{1}{\sqrt{3}}(|0\rangle + e^{-\texttt{i}\tilde{\delta}_1} |1\rangle + e^{-\texttt{i}\tilde{\delta}_2}|2\rangle)_{B_1},\\
	   &|\tilde{\varphi}_1\rangle_{B_1} = \frac{1}{\sqrt{3}}(|0\rangle + e^{\frac{\texttt{i}2\pi}{3}}  e^{-\texttt{i}\tilde{\delta}_1}|1\rangle
                                                                     + e^{\frac{\texttt{i}4\pi}{3}}  e^{-\texttt{i}\tilde{\delta}_2}|2\rangle)_{B_1},\\
	   &|\tilde{\varphi}_1\rangle_{B_1} = \frac{1}{\sqrt{3}}(|0\rangle + e^{\frac{\texttt{i}4\pi}{3}}  e^{-\texttt{i}\tilde{\delta}_1}|1\rangle
                                                                     + e^{\frac{\texttt{i}2\pi}{3}}  e^{-\texttt{i}\tilde{\delta}_2}|2\rangle)_{B_1}.
  \end{split}
\end{eqnarray}
After measurements, Alice (Bob) informs her (his) measurement result to Bob (Alice) and Charlie by classical channel.

Secondly, if Charlie wants to help Alice and Bob to transmit the states, he needs to make projective measurements on his particles $C_1$ and $C_2$ in the basis
$\{ |\overline{\varphi}_0\rangle, |\overline{\varphi}_1\rangle, |\overline{\varphi}_2\rangle\}$. Here
\begin{eqnarray}                         \label{6}
  \begin{split}
     &|\overline{\varphi}_0\rangle = \frac{1}{\sqrt{3}}(|0\rangle + |1\rangle + |2\rangle), \\
     &|\overline{\varphi}_1\rangle = \frac{1}{\sqrt{3}}(|0\rangle + e^{\frac{\texttt{i}2\pi}{3}} |1\rangle
                                                                  + e^{\frac{\texttt{i}4\pi}{3}} |2\rangle),\\
     &|\overline{\varphi}_2\rangle = \frac{1}{\sqrt{3}}(|0\rangle + e^{\frac{\texttt{i}4\pi}{3}} |1\rangle
                                                                  + e^{\frac{\texttt{i}2\pi}{3}} |2\rangle).
  \end{split}
  \end{eqnarray}
And then, Charlie informs his measurement results to Alice and Bob by classical channel.
Based on Equation \eqref{3}-Equation \eqref{6}, one can see that Equation \eqref{3} can be written as
\begin{eqnarray}                         \label{7}
	\begin{split}
		|\Psi\rangle = \frac{1}{3}\sum_{k,l,m,n=0}^2{|\overline{\varphi}_k\rangle_{C_2}
                                         \otimes |\varphi_l\rangle_{A_2}
                                         \otimes |\overline{\varphi}_m\rangle_{C_1}
                                         \otimes |\tilde{\varphi}_n\rangle_{B_1}
                                         \otimes |\tilde{x}_{m \bigoplus n}\rangle_{A_1}
                                         \otimes |x_{k \bigoplus l}\rangle_{B_2}},	
	\end{split}
\end{eqnarray}
where
\begin{eqnarray}                         \label{8}
  \begin{split}
    &|x_0\rangle_{B_2} = \frac{1}{\sqrt{3}}(|0\rangle + e^{\texttt{i}\delta_1}                               |1\rangle
                                                      + e^{\texttt{i}\delta_2}                               |2\rangle)_{B_2},\\
    &|x_1\rangle_{B_2} = \frac{1}{\sqrt{3}}(|0\rangle + e^{\frac{\texttt{i}4\pi}{3}}  e^{\texttt{i}\delta_1} |1\rangle
                                                      + e^{\frac{\texttt{i}2\pi}{3}}  e^{\texttt{i}\delta_2} |2\rangle)_{B_2},\\
    &|x_2\rangle_{B_2} = \frac{1}{\sqrt{3}}(|0\rangle + e^{\frac{\texttt{i}2\pi}{3}}  e^{\texttt{i}\delta_1} |1\rangle
                                                      + e^{\frac{\texttt{i}4\pi}{3}}  e^{\texttt{i}\delta_2} |2\rangle)_{B_2},
  \end{split}
\end{eqnarray}
\begin{eqnarray}                         \label{9}
 \begin{split}
	&|\tilde{x}_0\rangle_{A_1} = \frac{1}{\sqrt{3}}(|0\rangle + e^{\texttt{i}\tilde{\delta}_1}                                  |1\rangle
                                                            + e^{\texttt{i}\tilde{\delta}_2}                                  |2\rangle)_{A_1},\\
	&|\tilde{x}_1\rangle_{A_1} = \frac{1}{\sqrt{3}}(|0\rangle + e^{\frac{\texttt{i}4\pi}{3}}     e^{\texttt{i}\tilde{\delta}_1} |1\rangle
                                                            + e^{\frac{\texttt{i}2\pi}{3}}     e^{\texttt{i}\tilde{\delta}_2} |2\rangle)_{A_1},\\
	&|\tilde{x}_2\rangle_{A_1} = \frac{1}{\sqrt{3}}(|0\rangle + e^{\frac{\texttt{i}2\pi}{3}}     e^{\texttt{i}\tilde{\delta}_1} |1\rangle
                                                            + e^{\frac{\texttt{i}4\pi}{3}}     e^{\texttt{i}\tilde{\delta}_2} |2\rangle)_{A_1}.
 \end{split}
\end{eqnarray}

Thirdly,  Alice (Bob) performs the corresponding single-qutrit operations on particle $A_1$ ($B_2$) to reconstruct the target state
    $|\tilde{\chi}\rangle = \frac{1}{\sqrt{3}}(|0\rangle + e^{\texttt{i}\tilde{\delta}_1}|1\rangle + e^{\texttt{i}\tilde{\delta}_1}|2\rangle)$
           ($|\chi\rangle = \frac{1}{\sqrt{3}}(|0\rangle + e^{\texttt{i}\delta_1}|1\rangle         + e^{\texttt{i}\delta_2}|2\rangle)$)
from Bob (Alice), see Table \ref{Table1} for more detail.
\begin{eqnarray}                         \label{10}
 \begin{split}
	  U_0 |\tilde{x}_0\rangle_{A_1} = |\tilde{\chi}\rangle_{A_1}, \quad 
    U_1 |\tilde{x}_1\rangle_{A_1} = |\tilde{\chi}\rangle_{A_1}, \quad
    U_2 |\tilde{x}_2\rangle_{A_1} = |\tilde{\chi}\rangle_{A_1}.
 \end{split}
\end{eqnarray}
\begin{eqnarray}                         \label{11}
 \begin{split}
	  U_0 |x_0\rangle_{B_2} = |\chi\rangle_{B_2}, \quad 
    U_1 |x_1\rangle_{B_2} = |\chi\rangle_{B_2}, \quad 
    U_2 |x_2\rangle_{B_2} = |\chi\rangle_{B_2}.
 \end{split}
\end{eqnarray}
\begin{table}[htb]
  \centering
   \caption{The correspondences between the measurement results of Alice, Bob, and Charlie and the classical single-qutrit feed-forward operations. Here $U_{A_1}$ and $U_{B_2}$ are applied on particles $A_1$ and $B_2$, respectively.}
    \renewcommand{\arraystretch}{1.5}
     \setlength{\tabcolsep}{4mm}{
 \begin{tabular}{ccccc}
     \hline \hline
     $(M_{A_2},M_{C_2})$ & $M_{B_1}$ & $M_{C_1}$ & $U_{A_1}$  & $U_{B_2}$\\ \hline

                                                                     & $|\tilde{\varphi}_0\rangle_{B_1}$    &   $|\overline{\varphi}_0\rangle_{C_1}$  &  $U_0$  &  $U_0$\\
                                                                     & $|\tilde{\varphi}_1\rangle_{B_1}$    &   $|\overline{\varphi}_0\rangle_{C_1}$  &  $U_0$  &  $U_1$ \\
                                                                     & $|\tilde{\varphi}_2\rangle_{B_1}$    &   $|\overline{\varphi}_0\rangle_{C_1}$  &  $U_0$  &  $U_2$\\

  $(|\varphi_0\rangle_{A_2}, |\overline{\varphi}_0\rangle_{C_2})$    & $|\tilde{\varphi}_0\rangle_{B_1}$    &   $|\overline{\varphi}_1\rangle_{C_1}$  & $U_0$   &  $U_1$ \\
  $(|\varphi_1\rangle_{A_2}, |\overline{\varphi}_2\rangle_{C_2})$    & $|\tilde{\varphi}_1\rangle_{B_1}$    &   $|\overline{\varphi}_1\rangle_{C_1}$  & $U_0$   &  $U_2$ \\
  $(|\varphi_2\rangle_{A_2}, |\overline{\varphi}_1\rangle_{C_2})$    & $|\tilde{\varphi}_2\rangle_{B_1}$    &   $|\overline{\varphi}_1\rangle_{C_1}$  & $U_0$   &  $U_0$ \\

                                                                     & $|\tilde{\varphi}_0\rangle_{B_1}$    &   $|\overline{\varphi}_2\rangle_{C_1}$  & $U_0$   &  $U_2$ \\
                                                                     & $|\tilde{\varphi}_1\rangle_{B_1}$    &   $|\overline{\varphi}_2\rangle_{C_1}$  & $U_0$   &  $U_0$ \\
                                                                     & $|\tilde{\varphi}_2\rangle_{B_1}$    &   $|\overline{\varphi}_2\rangle_{C_1}$  & $U_0$   &  $U_1$ \\ \hline

                                                                     & $|\tilde{\varphi}_0\rangle_{B_1}$    &   $|\overline{\varphi}_0\rangle_{C_1}$  &  $U_1$  &  $U_0$ \\
                                                                     & $|\tilde{\varphi}_1\rangle_{B_1}$    &   $|\overline{\varphi}_0\rangle_{C_1}$  &  $U_1$  &  $U_1$ \\
                                                                     & $|\tilde{\varphi}_2\rangle_{B_1}$    &   $|\overline{\varphi}_0\rangle_{C_1}$  &  $U_1$  &  $U_2$ \\

$(|\varphi_0\rangle_{A_2},  |\overline{\varphi}_1\rangle_{C_2})$     & $|\tilde{\varphi}_0\rangle_{B_1}$    &   $|\overline{\varphi}_1\rangle_{C_1}$  &  $U_1$  &  $U_1$ \\
$(|\varphi_1\rangle_{A_2},  |\overline{\varphi}_0\rangle_{C_2})$     & $|\tilde{\varphi}_1\rangle_{B_1}$    &   $|\overline{\varphi}_1\rangle_{C_1}$  &  $U_1$  &  $U_2$ \\
$(|\varphi_2\rangle_{A_2},  |\overline{\varphi}_2\rangle_{C_2})$     & $|\tilde{\varphi}_2\rangle_{B_1}$    &   $|\overline{\varphi}_1\rangle_{C_1}$  &  $U_1$  &  $U_0$ \\

                                                                     & $|\tilde{\varphi}_0\rangle_{B_1}$    &   $|\overline{\varphi}_2\rangle_{C_1}$  &  $U_1$  &  $U_2$ \\
                                                                     & $|\tilde{\varphi}_1\rangle_{B_1}$    &   $|\overline{\varphi}_2\rangle_{C_1}$  &  $U_1$  &  $U_0$ \\
                                                                     & $|\tilde{\varphi}_2\rangle_{B_1}$    &   $|\overline{\varphi}_2\rangle_{C_1}$  &  $U_1$  &  $U_1$ \\ \hline

                                                                     & $|\tilde{\varphi}_0\rangle_{B_1}$    &   $|\overline{\varphi}_0\rangle_{C_1}$  &  $U_2$  &  $U_0$ \\
                                                                     & $|\tilde{\varphi}_1\rangle_{B_1}$    &   $|\overline{\varphi}_0\rangle_{C_1}$  &  $U_2$  &  $U_1$ \\
                                                                     & $|\tilde{\varphi}_2\rangle_{B_1}$    &   $|\overline{\varphi}_0\rangle_{C_1}$  &  $U_2$  &  $U_2$ \\

$(|\varphi_0\rangle_{A_2},  |\overline{\varphi}_2\rangle_{C_2})$     & $|\tilde{\varphi}_0\rangle_{B_1}$    &   $|\overline{\varphi}_1\rangle_{C_1}$  &  $U_2$  &  $U_1$ \\
$(|\varphi_1\rangle_{A_2},  |\overline{\varphi}_1\rangle_{C_2})$     & $|\tilde{\varphi}_1\rangle_{B_1}$    &   $|\overline{\varphi}_1\rangle_{C_1}$  &  $U_2$  &  $U_2$ \\
$(|\varphi_2\rangle_{A_2},  |\overline{\varphi}_0\rangle_{C_2})$     & $|\tilde{\varphi}_2\rangle_{B_1}$    &   $|\overline{\varphi}_1\rangle_{C_1}$  &  $U_2$  &  $U_0$ \\

                                                                     & $|\tilde{\varphi}_0\rangle_{B_1}$    &   $|\overline{\varphi}_2\rangle_{C_1}$  &  $U_2$  &  $U_2$ \\
                                                                     & $|\tilde{\varphi}_1\rangle_{B_1}$    &   $|\overline{\varphi}_2\rangle_{C_1}$  &  $U_2$  &  $U_0$ \\
                                                                     & $|\tilde{\varphi}_2\rangle_{B_1}$    &   $|\overline{\varphi}_2\rangle_{C_1}$  &  $U_2$  &  $U_1$ \\ \hline \hline
 \end{tabular}}\label{Table1}
 \end{table}
Here
\begin{eqnarray}                         \label{12}
	U_0= \left(
	\begin{array}{ccc}
	1 & 0 & 0\\
	0 & 1 & 0\\
	0 & 0 & 1
	\end{array}\right),
\end{eqnarray}
\begin{eqnarray}                         \label{13}
	U_1= \left(
	\begin{array}{ccc}
	1 & 0                     & 0\\
	0 & e^{\texttt{i}2\pi/3}  & 0\\
	0 & 0                     & e^{\texttt{i}4\pi /3}
	\end{array}\right),
\end{eqnarray}
\begin{eqnarray}                         \label{14}
	U_2= \left(
	\begin{array}{ccc}
	1 & 0                      & 0\\
	0 & e^{\texttt{i}4\pi /3}  & 0\\
	0 & 0                      & e^{\texttt{i}2\pi /3}
	\end{array}\right).
\end{eqnarray}

For instance, if the measurement result of Alice is $|\varphi_1\rangle_{A_2} $,
Bob is $|\tilde{\varphi}_1\rangle_{B_1}$, based on Equation \eqref{7}, one can see that the composite system composed of particles $A_1$, $A_2$, $B_1$, $B_2$, $C_1$ and $C_2$ will collapse into the state
\begin{eqnarray}                         \label{15}
  \begin{split}
    |\Psi'\rangle =   \; &\frac{1}{9}(|\tilde{x}_2\rangle_{A_1}  |\overline{\varphi}_0\rangle_{C_1}
                                    + |\tilde{x}_0\rangle_{A_1}  |\overline{\varphi}_1\rangle_{C_1}
                                    + |\tilde{x}_1\rangle_{A_1}  |\overline{\varphi}_2\rangle_{C_1})\\ & \otimes
                                     (|x_1\rangle_{B_2}          |\overline{\varphi}_0\rangle_{C_2}
                                    + |x_2\rangle_{B_2}          |\overline{\varphi}_1\rangle_{C_2}
                                    + |x_0\rangle_{B_2}          |\overline{\varphi}_2\rangle_{C_2}).
  \end{split}
\end{eqnarray}
We assume that Charlie would like to help Alice and Bob to complete the BCRSP, and the measurement results of particles $C_1$ and $C_2$ are  $|\overline{\varphi}_0\rangle_{C_1}$ and $|\overline{\varphi}_1\rangle_{C_2}$, respectively. Then  $|\Psi^{'}\rangle$  will project into the state
\begin{eqnarray}                         \label{16}
  \begin{split}
    |\Psi''\rangle =    \frac{1}{27}(|0\rangle + e^{\frac{\texttt{i}4\pi}{3}} e^{\texttt{i}\tilde\delta_1} |1\rangle
                                               + e^{\frac{\texttt{i}2\pi}{3}} e^{\texttt{i}\tilde\delta_2} |2\rangle)_{A_1} \otimes
                                    (|0\rangle + e^{\frac{\texttt{i}2\pi}{3}} e^{\texttt{i}{\delta}_1}     |1\rangle
                                               + e^{\frac{\texttt{i}4\pi}{3}} e^{\texttt{i}{\delta}_2}     |2\rangle)_{B_2}.
  \end{split}
\end{eqnarray}
Subsequently, Alice performs operation $U_1$ on her particle $A_1$ to obtain the desired state $|\tilde{\chi}\rangle$ from Bob.  Bob applies operation $U_2$ on his particle $B_2$ to obtain the desired state $|\chi\rangle$ from Alice.

\subsection{Single-particle four-dimensional state}\label{sec22}

Suppose that Alice and Bob wish to transmit their known single-qudit (four-dimensional) normalization state $|\omega\rangle_A$ and $|\tilde{\omega}\rangle_B$ to each other under the control of the supervisor Charlie, simultaneously.  Here
\begin{eqnarray}                         \label{17}
  \begin{split}
    |\omega\rangle_A = \frac{1}{2}(|0\rangle + e^{\texttt{i}\eta_1}|1\rangle
                                             + e^{\texttt{i}\eta_2}|2\rangle
                                             + e^{\texttt{i}\eta_3}|3\rangle)_A,
  \end{split}
\end{eqnarray}
\begin{eqnarray}                         \label{18}
	\begin{split}
		|\tilde{\omega}\rangle_B = \frac{1}{2}(|0\rangle + e^{\texttt{i}\tilde{\eta}_1} |1\rangle
                                                     + e^{\texttt{i}\tilde{\eta}_2} |2\rangle
                                                     + e^{\texttt{i}\tilde{\eta}_3} |3\rangle)_B.
	\end{split}
\end{eqnarray}
The real parameters $\eta_1$, $\eta_2$ and $\eta_3$ are only known by Alice. Real parameters $\tilde{\eta}_1$, $\tilde{\eta}_2$ and $\tilde{\eta}_2$ are only known by Bob.
To complete such task, we take the following channel
\begin{eqnarray}                        \label{19}
	\begin{split}
|\Phi\rangle = \frac{1}{2} (|000\rangle + |111\rangle + |222\rangle + |333\rangle)_{A_1B_1C_1}\otimes
		           \frac{1}{2} (|000\rangle + |111\rangle + |222\rangle + |333\rangle)_{A_2B_2C_2}.
	\end{split}
\end{eqnarray}
Alice possesses particles $A_1$ and $A_2$, Bob holds particles $B_1$ and $B_2$, and Charlie owns particles $C_1$ and $C_2$.

In the first step, Alice makes a single-particle projection measurement on particle $A_2$. At the same time, Bob makes a single-particle projection measurement on particle $B_1$. The measuring basis of $A_2$ is chosen as
\begin{eqnarray}                         \label{20}
	\begin{split}
     &|\upsilon_0\rangle_{A_2} = \frac{1}{2}(|0\rangle
                                          + e^{-\texttt{i}\eta_1}|1\rangle
                                          + e^{-\texttt{i}\eta_2}|2\rangle
                                          + e^{-\texttt{i}\eta_3}|3\rangle)_{A_2},\\
     &|\upsilon_1\rangle_{A_2} = \frac{1}{2}(|0\rangle
                                          + e^{\frac{\texttt{i}\pi}{2}}   e^{-\texttt{i}\eta_1}|1\rangle
                                          + e^{\texttt{i}\pi}             e^{-\texttt{i}\eta_2}|2\rangle
                                          + e^{\frac{\texttt{i}3\pi}{2}}  e^{-\texttt{i}\eta_3}|3\rangle)_{A_2},\\
     &|\upsilon_2\rangle_{A_2} = \frac{1}{2}(|0\rangle
                                          + e^{\texttt{i}\pi}             e^{-\texttt{i}\eta_1}|1\rangle
                                          + e^{\texttt{i}2\pi}            e^{-\texttt{i}\eta_2}|2\rangle
                                          + e^{\texttt{i}\pi}             e^{-\texttt{i}\eta_3}|3\rangle)_{A_2},\\
     &|\upsilon_3\rangle_{A_2} = \frac{1}{2}(|0\rangle
                                          + e^{\frac{\texttt{i}3\pi}{2}}  e^{-\texttt{i}\eta_1}|1\rangle
                                          + e^{\texttt{i}\pi}             e^{-\texttt{i}\eta_2}|2\rangle
                                          + e^\frac{\texttt{i}\pi}{2}     e^{-\texttt{i}\eta_3}|3\rangle)_{A_2}.
	\end{split}
\end{eqnarray}
The measuring basis of $B_1$ is chosen as
\begin{eqnarray}                         \label{21}
	\begin{split}
&|\tilde{\upsilon}_0\rangle_{B_1} = \frac{1}{2}(|0\rangle + e^{-\texttt{i}\tilde{\eta}_1} |1\rangle
                                                          + e^{-\texttt{i}\tilde{\eta}_2} |2\rangle
                                                          + e^{-\texttt{i}\tilde{\eta}_3} |3\rangle)_{B_1},\\
&|\tilde{\upsilon}_1\rangle_{B_1} = \frac{1}{2}(|0\rangle + e^{\frac{\texttt{i}\pi}{2}}  e^{-\texttt{i}\tilde{\eta}_1}|1\rangle
                                                          + e^{\texttt{i}\pi}            e^{-\texttt{i}\tilde{\eta}_2}|2\rangle
                                                          + e^\frac{\texttt{i}3\pi}{2}   e^{-\texttt{i}\tilde{\eta}_3}|3\rangle)_{B_1},\\
&|\tilde{\upsilon}_2\rangle_{B_1} = \frac{1}{2}(|0\rangle + e^{\texttt{i}\pi}            e^{-\texttt{i}\tilde{\eta}_1}|1\rangle
                                                          + e^{\texttt{i}2\pi}           e^{-\texttt{i}\tilde{\eta}_2}|2\rangle
                                                          + e^{\texttt{i}\pi}            e^{-\texttt{i}\tilde{\eta}_3}|3\rangle)_{B_1},\\
&|\tilde{\upsilon}_3\rangle_{B_1} = \frac{1}{2}(|0\rangle + e^{\frac{\texttt{i}3\pi}{2}} e^{-\texttt{i}\tilde{\eta}_1}|1\rangle
                                                          + e^{\texttt{i}\pi}            e^{-\texttt{i}\tilde{\eta}_2}|2\rangle
                                                          + e^\frac{\texttt{i}\pi}{2}    e^{-\texttt{i}\tilde{\eta}_3}|3\rangle)_{B_1}.
	\end{split}
\end{eqnarray}
Then Alice and Bob publicly convey their measurement outcomes.

In the second step, after Charlie obtaining the results from Alice and Bob,
if he wishes to help them, he needs to measure his particles $C_1$ and $C_2$ in the orthogonal basis
$\{|\overline{\upsilon}_0\rangle, |\overline{\upsilon}_1\rangle, |\overline{\upsilon}_2\rangle,|\overline{\upsilon}_3\rangle\}$, respectively.
Here
\begin{eqnarray}                         \label{22}
  \begin{split}
       &|\overline{\upsilon}_0\rangle = \frac{1}{2}(|0\rangle + |1\rangle + |2\rangle + |3\rangle),\\
       &|\overline{\upsilon}_1\rangle = \frac{1}{2}(|0\rangle + e^{\frac{\texttt{i}\pi}{2}} |1\rangle
                                                              + e^{\texttt{i}\pi}           |2\rangle
                                                              + e^{\frac{\texttt{i}3\pi}{2}}|3\rangle),\\
       &|\overline{\upsilon}_2\rangle = \frac{1}{2}(|0\rangle + e^{\texttt{i}\pi}           |1\rangle
                                                              + e^{\texttt{i}2\pi}          |2\rangle
                                                              + e^{\texttt{i}\pi}           |3\rangle),\\
       &|\overline{\upsilon}_3\rangle = \frac{1}{2}(|0\rangle + e^{\frac{\texttt{i}3\pi}{2}}|1\rangle
                                                              + e^{\texttt{i}\pi}           |2\rangle
                                                              + e^\frac{\texttt{i}\pi}{2}   |3\rangle).
  \end{split}
\end{eqnarray}
Then Charlie tells the two others about his measurement results through the classical channel. Based on Equation \eqref{19}-Equation \eqref{22}, we can see that Equation \eqref{19} can be rewritten as
\begin{eqnarray}                         \label{23}
	\begin{split}
|\Phi\rangle = \frac{1}{4}\sum_{k,l,m,n=0}^3{|\overline{\upsilon}_k\rangle_{C_2}
                                     \otimes |\upsilon_l\rangle_{A_2}
                                     \otimes |\overline{\upsilon}_m\rangle_{C_1}
                                     \otimes |\tilde{\upsilon}_n\rangle _{B_1}
		                                 \otimes |\tilde{y}_{m \bigoplus n}\rangle_{A_1}
                                     \otimes |y_{k \bigoplus l}\rangle_{B_2}},	
	\end{split}
\end{eqnarray}
where
\begin{eqnarray}                         \label{24}
	\begin{split}
    &|y_0\rangle = \frac{1}{2}(|0\rangle + e^{\texttt{i}\eta_1}|1\rangle
                                         + e^{\texttt{i}\eta_2}|2\rangle
                                         + e^{\texttt{i}\eta_3}|3\rangle),\\
    &|y_1\rangle = \frac{1}{2}(|0\rangle + e^{\frac{\texttt{i}3\pi}{2}} e^{\texttt{i}\eta_1}|1\rangle
                                         + e^{\texttt{i}\pi}            e^{\texttt{i}\eta_2}|2\rangle
                                         + e^\frac{\texttt{i}\pi}{2}    e^{\texttt{i}\eta_3}|3\rangle),\\
    &|y_2\rangle = \frac{1}{2}(|0\rangle + e^{\texttt{i}\pi}            e^{\texttt{i}\eta_1}|1\rangle
                                         + e^{\texttt{i}2\pi}           e^{\texttt{i}\eta_2}|2\rangle
                                         + e^{\texttt{i}\pi}            e^{\texttt{i}\eta_3}|3\rangle),\\
    &|y_3\rangle = \frac{1}{2}(|0\rangle + e^{\frac{\texttt{i}\pi}{2}}  e^{\texttt{i}\eta_1}|1\rangle
                                         + e^{\texttt{i}\pi}            e^{\texttt{i}\eta_2}|2\rangle
                                         + e^\frac{\texttt{i}3\pi}{2}   e^{\texttt{i}\eta_3}|3\rangle),
	\end{split}
\end{eqnarray}
\begin{eqnarray}                         \label{25}
	\begin{split}
&|\tilde{y}_0\rangle = \frac{1}{2}(|0\rangle + e^{\texttt{i}\tilde{\eta}_1}                                 |1\rangle
                                             + e^{\texttt{i}\tilde{\eta}_2}                                 |2\rangle
                                             + e^{\texttt{i}\tilde{\eta}_3}                                 |3\rangle),\\
&|\tilde{y}_1\rangle = \frac{1}{2}(|0\rangle + e^{\frac{\texttt{i}3\pi}{2}}     e^{\texttt{i}\tilde{\eta}_1}|1\rangle
                                             + e^{\texttt{i}\pi}                e^{\texttt{i}\tilde{\eta}_2}|2\rangle
                                             + e^\frac{\texttt{i}\pi}{2}        e^{\texttt{i}\tilde{\eta}_3}|3\rangle),\\
&|\tilde{y}_2\rangle = \frac{1}{2}(|0\rangle + e^{\texttt{i}\pi}                e^{\texttt{i}\tilde{\eta}_1}|1\rangle
                                             + e^{\texttt{i}2\pi}               e^{\texttt{i}\tilde{\eta}_2}|2\rangle
                                             + e^{\texttt{i}\pi}                e^{\texttt{i}\tilde{\eta}_3}|3\rangle),\\
&|\tilde{y}_3\rangle = \frac{1}{2}(|0\rangle + e^{\frac{\texttt{i}\pi}{2}}      e^{\texttt{i}\tilde{\eta}_1}|1\rangle
                                             + e^{\texttt{i}\pi}                e^{\texttt{i}\tilde{\eta}_2}|2\rangle
                                             + e^\frac{\texttt{i}3\pi}{2}       e^{\texttt{i}\tilde{\eta}_3}|3\rangle).
	\end{split}
\end{eqnarray}

In the third step, after Alice and Bob receive the measurement results from Charlie, Alice can recover her original state
$|\tilde{\omega}\rangle = \frac{1}{2}(|0\rangle + e^{\texttt{i}\tilde{\eta}_1} |1\rangle
                                                + e^{\texttt{i}\tilde{\eta}_2} |2\rangle
                                                + e^{\texttt{i}\tilde{\eta}_3} |3\rangle)$
by performing the corresponding classical single-qudit feed-forward operations
\begin{eqnarray}                         \label{26}
 \begin{split}
  &U_0 |\tilde{y}_0\rangle_{A_1} = |\tilde{\omega}\rangle_{A_1}, \quad
   U_1 |\tilde{y}_1\rangle_{A_1} = |\tilde{\omega}\rangle_{A_1}, \\
  &U_2 |\tilde{y}_2\rangle_{A_1} = |\tilde{\omega}\rangle_{A_1}, \quad
   U_3 |\tilde{y}_3\rangle_{A_1} = |\tilde{\omega}\rangle_{A_1},
 \end{split}
\end{eqnarray}
on her remain photon $A_1$. Bob can recover his original state
$|\omega\rangle = \frac{1}{2} (|0\rangle + e^{\texttt{i}\eta_1}|1\rangle
                                         + e^{\texttt{i}\eta_2}|2\rangle
                                         + e^{\texttt{i}\eta_3}|3\rangle)$
by performing the corresponding classical single-qudit feed-forward operations
\begin{eqnarray}                        \label{27}
 \begin{split}
  &U_0 |y_0\rangle_{B_2} = |\omega\rangle_{B_2}, \quad
   U_1 |y_1\rangle_{B_2} = |\omega\rangle_{B_2}, \\
  &U_2 |y_2\rangle_{B_2} = |\omega\rangle_{B_2}, \quad
   U_3 |y_3\rangle_{B_3} = |\omega\rangle_{B_2},
 \end{split}
\end{eqnarray}
on his remain photon $B_2$.
Here
\begin{eqnarray}                        \label{28}
	U_0= \left(
	 \begin{array}{cccc}
	1 & 0 & 0 & 0\\
	0 & 1 & 0 & 0\\
	0 & 0 & 1 & 0\\
	0 & 0 & 0 & 1
   \end{array}\right),
\end{eqnarray}
\begin{eqnarray}                         \label{29}
	U_1= \left(
	 \begin{array}{cccc}
	1 & 0                    & 0                 & 0\\
	0 & e^{\texttt{i}\pi/2}  & 0                 & 0\\
	0 & 0                    & e^{\texttt{i}\pi} & 0\\
	0 & 0                    & 0                 & e^{\texttt{i}3\pi/2}
   \end{array}\right),
\end{eqnarray}
\begin{eqnarray}                         \label{30}
	U_2= \left(
	 \begin{array}{cccc}
	1 & 0                    & 0                  & 0\\
	0 & e^{\texttt{i}\pi}    & 0                  & 0\\
	0 & 0                    & 1                  & 0\\
	0 & 0                    & 0                  & e^{\texttt{i}\pi}
	\end{array}\right),
\end{eqnarray}
\begin{eqnarray}                         \label{31}
	U_3= \left(
	\begin{array}{cccc}
	1 & 0                    & 0                 & 0\\
	0 & e^{\texttt{i}3\pi/2} & 0                 & 0\\
	0 & 0                    & e^{\texttt{i}\pi} & 0\\
	0 & 0                    & 0                 & e^{\texttt{i}\pi/2}
	\end{array}\right).
\end{eqnarray}

For instance, when the measurement results of Alice and Bob are $|\upsilon_2\rangle_{A_2}$ and $|\tilde{\upsilon}_2\rangle_{B_1}$, the state described in Equation \eqref{23} will become
\begin{eqnarray}                         \label{32}
	\begin{split}
	|\Phi'\rangle  = \; &\frac{1}{16}(|\tilde{y_2}\rangle_{A_1}  |\overline{\upsilon}_0\rangle_{C_1}  +
                                      |\tilde{y_3}\rangle_{A_1}|\overline{\upsilon}_1\rangle_{C_1}  +
                                      |\tilde{y_0}\rangle_{A_1}|\overline{\upsilon}_2\rangle_{C_1}  +
                                      |\tilde{y_1}\rangle_{A_1}|\overline{\upsilon}_3\rangle_{C_1}) \\ & \otimes
                                            (|y_2\rangle_{B_2} |\overline{\upsilon}_0\rangle_{C_2}  +
                                             |y_3\rangle_{B_2} |\overline{\upsilon}_1\rangle_{C_2}  +
                                             |y_0\rangle_{B_2} |\overline{\upsilon}_2\rangle_{C_2}  +
                                             |y_1\rangle_{B_2} |\overline{\upsilon}_3\rangle_{C_2}).
	\end{split}
\end{eqnarray}

Nextly, if Charlie's measurement results are $|\overline{\upsilon}_0\rangle_{C_1}$ and $|\overline{\upsilon}_1\rangle_{C_2}$, $|\Phi'\rangle$ will project into the state
\begin{eqnarray}                         \label{33}
	\begin{split}
	|\Phi''\rangle  =\; &\frac{1}{64}(|0\rangle + e^{\texttt{i}\pi}                 e^{\texttt{i}\tilde{\eta}_1}|1\rangle
                                                 + e^{\texttt{i}2\pi}             e^{\texttt{i}\tilde{\eta}_2}|2\rangle
                                                 + e^{\texttt{i}\pi}              e^{\texttt{i}\tilde{\eta}_3}|3\rangle)_{A_1}\\ & \otimes
                                      (|0\rangle + e^{\frac{\texttt{i}\pi}{2}}    e^{\texttt{i}\eta_1}        |1\rangle
                                                 + e^{\texttt{i}\pi}              e^{\texttt{i}\eta_2}        |2\rangle
                                                 + e^\frac{\texttt{i}3\pi}{2}     e^{\texttt{i}\eta_3}        |3\rangle)_{B_2}.
	\end{split}
\end{eqnarray}
Lastly, Alice will perform $U_2$ on particle $A_1$ to reconstruct the original state $|\tilde{\omega}\rangle$ from Bob, and
Bob will perform $U_3$ on particle $B_1$ to reconstruct the original state $|\omega\rangle$ from Alice.

\subsection{Single-particle $N$-dimensional state}\label{sec23}

Our scheme can be extended to implement BCRSP in $N$-dimensional system. Suppose Alice wants to help Bob remotely prepare a known normalization single-quNit ($N$-dimensional) state
\begin{eqnarray}                         \label{34}
  \begin{split}
    |\nu\rangle_A = \frac{1}{\sqrt{N}} (|0\rangle + e^{\texttt{i}\vartheta_1}    |1\rangle
                                                  + e^{\texttt{i}\vartheta_2}    |2\rangle
                                                  + \cdots
                                                  + e^{\texttt{i}\vartheta_{N-1}}|N-1\rangle)_A.
  \end{split}
\end{eqnarray}
At the same time, Bob wants to help Alice remotely prepare another known single-quNit state
\begin{eqnarray}                         \label{35}
	\begin{split}
		|\tilde{\nu}\rangle_B = \frac{1}{\sqrt{N}}(|0\rangle + e^{\texttt{i}\tilde{\vartheta}_1}     |1\rangle
                                                         + e^{\texttt{i}\tilde{\vartheta}_2}     |2\rangle
                                                         + \cdots
                                                         + e^{\texttt{i}\tilde{\vartheta}_{N-1}} |N-1\rangle)_B.
	\end{split}
\end{eqnarray}
Charlie plays a role of supervisor. The quantum channel linking  Alice, Bob and Charlie is given by
\begin{eqnarray}                         \label{36}
	\begin{split}
    |\Omega\rangle =\; &\frac{1}{\sqrt{N}} (|000\rangle  + |111\rangle   + |222\rangle + \cdots  + |N-1,N-1,N-1\rangle)_{A_1B_1C_1}\\&\otimes
		                    \frac{1}{\sqrt{N}} (|000\rangle  + |111\rangle   + |222\rangle + \cdots  + |N-1,N-1,N-1\rangle)_{A_2B_2C_2},
	\end{split}
\end{eqnarray}
where particles $A_1$ and $A_2$ hold by Alice, particles $B_1$ and $B_2$ owned by Bob, and particles $C_1$ and $C_2$ belong to Charlie.
Our scheme can be completed by the following steps.

Step 1:  Alice performs a projective measurement on particle $A_2$ in the basis $\{|\tau_l\rangle \}$. At the same time, Bob  performs a projective measurement on particle $B_1$ in the basis $\{|\tilde{\tau}_n\rangle \}$. Here
\begin{eqnarray}                         \label{37}
	\begin{split}
	  |\tau_l\rangle = \frac{1}{\sqrt{N}} \sum_j{e}^{\texttt{i}2\pi jl/N}{e}^{-\texttt{i}\vartheta_j}|j\rangle,
	\end{split}
\end{eqnarray}
\begin{eqnarray}                         \label{38}
	\begin{split}
    |\tilde{\tau}_n\rangle = \frac{1}{\sqrt{N}} \sum_j{e}^{\texttt{i}2\pi jn/N}{e}^{-\texttt{i}\tilde{\vartheta}_j}|j\rangle,
	\end{split}
\end{eqnarray}
where $j,n,l=0,\;1,\;2,\cdots,\;N-1$, $\vartheta_0=0$, and $\tilde{\vartheta}_0=0$.
After the measurements, Alice and Bob inform their measurement results to each other and Charlie.

Step 2:  After Charlie receives the messages, he will measure particles $C_1$ and $C_2$ in the orthogonal basis $\{|\overline{\tau}_k\rangle \}$. Here
\begin{eqnarray}                         \label{39}
	\begin{split}
	  |\overline{\tau}_k\rangle = \frac{1}{\sqrt{N}} \sum_j{e}^{\texttt{i}2\pi jk/N}|j\rangle, \;\; j,k=0,\;1,\;2,\;\cdots,\;N-1.
	\end{split}
\end{eqnarray}
In the basis of $\{|\tau_l\rangle_{A_2}, |\tilde{\tau}_n\rangle_{B_1}, |\overline{\tau}_m\rangle_{C_1}, \overline{\tau}_k\rangle_{C_2} \}$,
Equation \eqref{36} can be rewritten as
\begin{eqnarray}                         \label{40}
	\begin{split}
		|\Omega\rangle = \frac{1}{N}\sum_{k,l,m,n=0}^{N-1}{|\overline{\tau}_k\rangle_{C_2}
                                               \otimes |\tau_l\rangle_{A_2}
                                               \otimes |\overline{\tau}_m\rangle_{C_1}
                                               \otimes |\tilde{\tau}_n\rangle_{B_1}
		                                           \otimes |\tilde{z}_{m\bigoplus n}\rangle_{A_1}}
                                               \otimes |z_{k\bigoplus l}\rangle_{B_2},	
	\end{split}
\end{eqnarray}
where
\begin{eqnarray}                         \label{41}
	\begin{split}
		|z_p\rangle = \frac{1}{\sqrt{N}} \sum_j{e}^{\texttt{i}2\pi jp/N}e^{\texttt{i}\vartheta_j}|j\rangle,
	\end{split}
\end{eqnarray}
\begin{eqnarray}                         \label{42}
	\begin{split}
		|\tilde{z}_q\rangle = \frac{1}{\sqrt{N}} \sum_j{e}^{\texttt{i}2\pi jq/N}e^{\texttt{i}\tilde{\vartheta_j}}|j\rangle,
	\end{split}
\end{eqnarray}
and $j,p,q=0,\;1,\;2,\;\cdots,\;N-1$. Then Charlie informs his measurements to Alice and Bob.

Step 3: Based on the measurement results of Alice, Bob, and Charlie, Alice and Bob will perform the corresponding classical single-particle operations $U_k$ to recover their original states, respectively. In detail
\begin{eqnarray}                         \label{43}
	\begin{split}
		&U_k|\tilde{z_k}\rangle_{A_1} = |\tilde{\nu}\rangle,
            &U_k|z_k\rangle_{B_2} = |\nu\rangle.
	\end{split}
\end{eqnarray}
where
\begin{eqnarray}                         \label{44}
	\begin{split}
		U_k = \sum_j{e}^{\texttt{i}2\pi jk/N}|j\rangle\langle j|, \;\; j,k=0,\;1,\;2,\;\cdots,\;N-1.
	\end{split}
\end{eqnarray}

\section{Discussion and Conclusion}      \label{sec3}

In experiment, the system is inevitable suffer from decoherence due to the inherent noise in quantum channel, which may degrade the performance of the protocol.\upcite{noise1,noise2,noise3} In the following we take the scenario in four-dimensional system as a representative example to evaluate the performance of our schemes with various noises.

\subsection{BCRSP in qudit-flip noisy channels} \label{sec31}

In our scheme shown in Section \ref{sec22}, the pre-shared state $|\Phi_1\rangle = \frac{1}{2}(|000\rangle+|111\rangle+|222\rangle+|333\rangle)_{A_1B_1C_1}$ ($|\Phi_2\rangle = \frac{1}{2}(|000\rangle+|111\rangle+|222\rangle+|333\rangle)_{A_2B_2C_2}$) is prepared by Alice (Bob) and then particles $B_1$ and $C_1$ ($A_2$ and $C_2$) are distributed to
Bob and Charlie (Alice and Charlie), respectively. During the particles distribution, if particles $A_2, B_1, C_1$ and $C_2$ suffer from qudit-flip noise,
the density matrix of qudits $A_1$, $A_2$, $B_1$, $B_2$, $C_1,$ and $C_2$ can be written as
\begin{eqnarray}                         \label{45}
	\begin{split}
		\varepsilon_1(\rho^{F})  =\; & \sum_{\substack{l_1,l_2,l_3,l_4=0}}^{3}(E_{l_1})_{B_1} (E_{l_2})_{C_1} (E_{l_3})_{A_2}(E_{l_4})_{C_2}\rho((E_{l_1})_{B_1} (E_{l_2})_{C_1} (E_{l_3})_{A_2}(E_{l_4})_{C_2})^\dagger \\
	                           =   & \frac{1}{16}\sum_{\substack{l_1,l_2,l_3,l_4,\\j_1,j_2,j_3,j_4=0}}^{3} \gamma_{l_1}^2 \gamma_{l_2}^2 \gamma_{l_3}^2 \gamma_{l_4}^2 |j_1,j_1\oplus l_1,j_1\oplus l_2,j_2\oplus l_3,j_2,j_2\oplus l_4\rangle \\ & \otimes
                                   \langle j_3,j_3\oplus l_1,j_3\oplus l_2,j_4\oplus l_3,j_4,j_4\oplus l_4|.
   \end{split}
\end{eqnarray}
Here density matrix $\rho = |\Phi\rangle \langle\Phi|$, and $|\Phi\rangle$ is described in Equation \eqref{19}.
$E_l$ is the Kraus operator of the qudit-flip noise, and it can be expressed as
\begin{eqnarray}                         \label{46}
	\begin{split}
		E_l = \gamma_l\sum_{j=0}^{3}|j\oplus l\rangle\langle j|, \;\; l=0,\;1,\;2,\;3.
	\end{split}
\end{eqnarray}
Here $\gamma_l$ represents the probability of error due to the presence of qudit-flip noise\upcite{noise}. The relations between $\gamma_l$ and the noise factor $\gamma$ are given by
\begin{eqnarray}                         \label{47}
	\begin{split}
		 \gamma_0 = \sqrt{1-\frac{3\gamma}{4}}, \;\;
     \gamma_1 = \gamma_2 = \gamma_3 = \sqrt{\frac{\gamma}{4}}.
	\end{split}
\end{eqnarray}

After the BCRSP described in Section \ref{sec22} is completed, Alice recovers the renewed quantum state
\begin{eqnarray}                         \label{48}
	\begin{split}
		|\tilde{\omega}^{F}\rangle = \frac{1}{2}\sum_{l_1,l_2,j=0}^{3} \gamma_{l_1} \gamma_{l_2} e^{\texttt{i}\tilde{\eta}_{j\oplus l_1}}|j\oplus l_1\rangle,
   \end{split}
\end{eqnarray}
on her qudit $A_1$. Bob recovers the renewed quantum state
\begin{eqnarray}                         \label{49}
	\begin{split}
		|\omega^{F}\rangle = \frac{1}{2}\sum_{l_3,l_4,j=0}^{3} \gamma_{l_3} \gamma_{l_4} e^{\texttt{i}\eta_{j\oplus l_3}}|j\oplus l_3\rangle,
   \end{split}
\end{eqnarray}
on his qudit $B_2$.
Therefore, the fidelity between the ideal outcome state $|\tilde{\omega}\rangle$ described in Equation \eqref{17} and the real outcome state $|\tilde{\omega}^{F}\rangle$
described in  Equation \eqref{48} can be calculated as
\begin{eqnarray}                         \label{50}
	\begin{split}
		 F_1(|\tilde{\omega}\rangle,\tilde{\rho}^{F}) \; &  = \sqrt{\langle\tilde{\omega}|\tilde{\rho}^{F}|\tilde{\omega}\rangle} \\
                                                     &  = \frac{1}{4} \sqrt{\sum_{l_1,l_2=0}^{3} \gamma_{l_1}^2 \gamma_{l_2}^2 \left\lvert\sum_{j_1,j_2,j,j^{'}=0}^{3} e^{\texttt{i}(\tilde{\eta}_{{j_1}\oplus l_1} + \tilde{\eta}_j)} e^{-\texttt{i}(\tilde{\eta}_{{j_3}\oplus l_1} +\tilde{\eta}_{j^{'}})}\right\rvert^2},
  \end{split}
\end{eqnarray}
where
\begin{eqnarray}                         \label{51}
	\begin{split}
		\tilde{\rho}^{F} \; & = |\tilde{\omega}^{F}\rangle \langle\tilde{\omega}^{F}|.
   \end{split}
\end{eqnarray}
The same result can also be obtained in Bob's site.
We can see that when the parameters of single-qudit state $|\tilde{\omega}\rangle = \frac{1}{2}(|0\rangle+|1\rangle+|2\rangle+|3\rangle)$.
The unity fidelity of BCRSP scheme in qudit-flip noisy environment  can be achieved.

\subsection{BCRSP in dephasing noisy channels} \label{sec32}

The Kraus operator of the dephasing noisy channel are given by
\begin{eqnarray}                         \label{53}
	\begin{split}
		E_l = \sum_{j=0}^{3}\gamma_{lj}|j\rangle\langle j|, \;\; l=0,\;1,\;2,\;3,
	\end{split}
\end{eqnarray}
where
\begin{eqnarray}                         \label{54}
	\begin{split}
		& \gamma_{00} = 1, \\
    & \gamma_{01} = \gamma_{02} = \gamma_{03} = \sqrt{1-\gamma}, \\
    & \gamma_{11} = \gamma_{22} = \gamma_{33} = \sqrt{\gamma}, \\
    & \gamma_{10} = \gamma_{12} = \gamma_{13} = \gamma_{20} = \gamma_{21} =\gamma_{23} =\gamma_{30} = \gamma_{31} = \gamma_{32} = 0.
	\end{split}
\end{eqnarray}
Here $0 \leq\gamma\leq 1$ represents the decoherent rate. As shown in Section \ref{sec22}, if particles $A_2, B_1, C_1$ and $C_2$ undergo the dephasing noise channel, then the renewed state of qudits $A_1, A_2, B_1, B_2, C_1, C_2$ can be expressed as
\begin{eqnarray}                         \label{55}
	\begin{split}
		\varepsilon_2(\rho^{P}) = \;
	                         &\frac{1}{16}\sum_{\substack{l_1,l_2,l_3,l_4,\\j_1,j_2,j_3,j_4,j_5,j_6,j_7,j_8=0}}^{3} \gamma_{l_1j_1} \gamma_{l_2j_2} \gamma_{l_3j_3} \gamma_{l_4j_4} \gamma_{l_1j_5} \gamma_{l_2j_6} \gamma_{l_3j_7} \gamma_{l_4j_8} |j_1,j_1,j_1,j_2,j_2,j_2\rangle \\&\otimes
                              \langle j_3,j_3,j_3,j_4,j_4,j_4|.
   \end{split}
\end{eqnarray}

After the BCRSP is accomplished, Alice recreates the renewed quantum state
\begin{eqnarray}                         \label{56}
	\begin{split}
		|\tilde{\omega}^{P}\rangle = \frac{1}{2}\sum_{\substack{l_1,l_2,j_1,j_2,j=0}}^{3} \gamma_{l_1j_1} \gamma_{l_2j_2} e^{\texttt{i}\tilde{\eta}_j}|j\rangle,
   \end{split}
\end{eqnarray}
on her particle $A_1$.
The fidelity between $|\tilde{\omega}\rangle$ described in Equation \eqref{17} and $|\tilde{\omega}^{P}\rangle$ described in Equation \eqref{56} can be calculated as
\begin{eqnarray}                         \label{57}
	\begin{split}
		 F_2(|\tilde{\omega}\rangle,\tilde{\rho}^{P}) \; & = \sqrt{\langle\tilde{\omega}|\tilde{\rho}^{P}|\tilde{\omega}\rangle} \\
                                                     & = \frac{1}{16} \sqrt{\sum_{l_1,l_2=0}^{3} \left\lvert \sum_{j_1,j_2=0}^{3} \gamma_{l_1j_1} \gamma_{l_2j_2} \right\rvert^2} \\
                                                     & = \frac{1}{16} \sqrt{[1+6\sqrt{1-\gamma}+9(1-\gamma)]^2 + 6\gamma(1+3\sqrt{1-\gamma})^2 + 9\gamma^2}.
   \end{split}
\end{eqnarray}

\subsection{BCRSP in qudit-phase-flip noisy channels} \label{sec33}

The Kraus operator of qudit-phase-flip noisy channel are given by
\begin{eqnarray}                         \label{58}
	\begin{split}
		E_{l_1l_2} = \gamma_{l_1l_2}\sum_{j=0}^{3} e^{\frac{\texttt{i}\pi}{2}jl_1}|j\oplus l_2\rangle\langle j|,\;\;l_1,l_2=0,\;1,2,\;3,
   \end{split}
\end{eqnarray}
where
\begin{eqnarray}                         \label{59}
	\begin{split}
		& \gamma_{00} = \sqrt{1-\frac{3\gamma}{4}},    \\
    & \gamma_{01} = \gamma_{02} = \gamma_{03} = 0, \\
    & \gamma_{10} = \gamma_{20} = \gamma_{30} = 0, \\
    & \gamma_{s_1s_2} = \sqrt{\frac{\gamma}{12}}, \;\; s_1,s_2=0,\;1,2,\;3.
	\end{split}
\end{eqnarray}
Here $\gamma$ indicates the decoherent rate. Then the qudit-phase-flip noisy environment will make the state of particles $A_1, A_2, B_1, B_2, C_1$ and $C_2$ become
\begin{eqnarray}                         \label{60}
	\begin{split}
		\varepsilon_3(\rho^{PF})  = \;
	                         &\frac{1}{16}\sum_{\substack{l_1,l_2,l_3,l_4,l_5,l_6,l_7,l_8,\\j_1,j_2,j_3,j_4=0}}^{3} \gamma_{l_1l_2}^2 \gamma_{l_3l_4}^2 \gamma_{l_5l_6}^2 \gamma_{l_7l_8}^2 e^{\frac{\texttt{i}\pi}{2}j_1l_1} e^{\frac{\texttt{i}\pi}{2}j_1l_3} e^{\frac{\texttt{i}\pi}{2}j_2l_5} e^{\frac{\texttt{i}\pi}{2}j_2l_7}|j_1,j_1 \oplus l_2,j_1 \oplus l_4,j_2 \oplus l_6, \\ & j_2 \oplus l_8\rangle
                              e^{\frac{-\texttt{i}\pi}{2}j_3l_1} e^{\frac{-\texttt{i}\pi}{2}j_3l_3} e^{\frac{-\texttt{i}\pi}{2}j_4l_5} e^{\frac{-\texttt{i}\pi}{2}j_4l_7} \langle j_3,j_3 \oplus l_2,j_3 \oplus l_4,j_4 \oplus l_6,j_4,j_4 \oplus l_8|.
   \end{split}
\end{eqnarray}

After the BCRSP described in Section \ref{sec22} is accomplished,  Alice recovers the renewed quantum state
\begin{eqnarray}                         \label{61}
	\begin{split}
		|\tilde{\omega}^{PF}\rangle = \frac{1}{2}\sum_{\substack{l_1,l_2,l_3,l_4=0}}^{3} \gamma_{l_1l_2} \gamma_{l_3l_4} e^{\frac{\texttt{i}\pi(l_1+l_3)j}{2}} e^{\texttt{i}\tilde{\eta}_j}|j\rangle,
   \end{split}
\end{eqnarray}
on her qudit $A_1$.
Then the fidelity between $|\tilde{\omega}\rangle$ described in Equation \eqref{17} and $|\tilde{\omega}^{PF}\rangle$ described in Equation \eqref{61} can be calculated as
\begin{eqnarray}                         \label{62}
	\begin{split}
		 F_3(|\tilde{\omega}\rangle,\tilde{\rho}^{PF}) \; & = \sqrt{\langle\tilde{\omega}|\tilde{\rho}^{PF}|\tilde{\omega}\rangle} \\
                                                      & = \frac{1}{4} \sqrt{\sum_{l_1,l_2,l_3,l_4-0}^{3} \gamma_{l_1l_2}^2 \gamma_{l_3l_4}^2 \left\lvert \sum_{j_1,j_2}^{3} e^{\frac{\texttt{i}\pi(l_1+l_3)j_1}{2}} e^{-\frac{\texttt{i}\pi(l_1+l_3)j_2}{2}} e^{\texttt{i}\tilde{\eta}_{j_1}} e^{-\texttt{i}\tilde{\eta}_{j_2}} \right\rvert^2}.
   \end{split}
\end{eqnarray}
In particular, for the equatorial state $|\tilde{\omega}\rangle = \frac{1}{2}(|0\rangle+|1\rangle+|2\rangle+|3\rangle)$, the fidelity $1-\frac{3\gamma}{4}$ can be achieved.

\subsection{The feasibilities of our BCRSPs} \label{sec34}

\begin{figure}[htbp]
  \centering
  \includegraphics[width=10 cm]{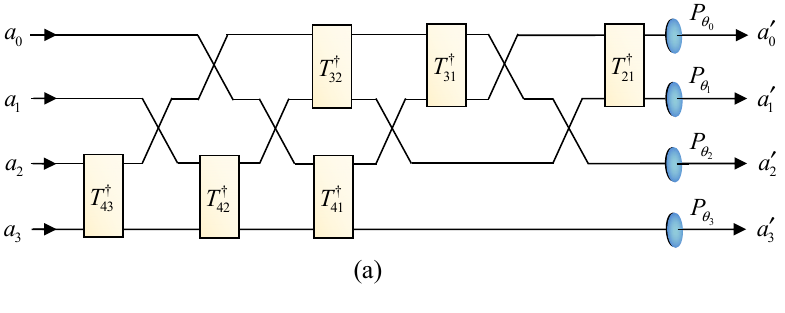}\\
  \includegraphics[width=10 cm]{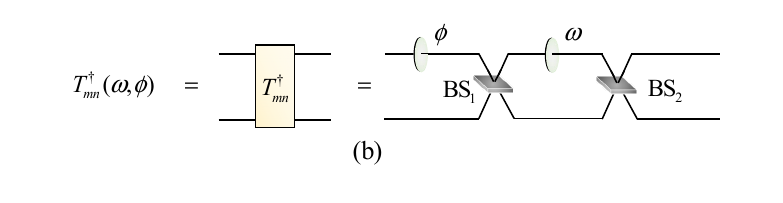}
  \caption{
    (a) A schematic diagram of the transformation described in Equation \eqref{65}. (b) A schematic diagram of a variable beam splitter $T_{mn}^\dag(\omega,\varphi)$. Phase shifters $P_{\theta_0}=P_{\theta_1}=e^{\texttt{i}\arctan(-\frac{1}{3})}$, $P_{\theta_2}=e^{\frac{\texttt{i}\pi}{4}}$, and $P_{\theta_3}=e^{\frac{\texttt{i}\pi}{2}}$.}
  \label{Fig.2}
\end{figure}

The generalized $N$-dimensional GHZ state $|\text{GHZ}\rangle = \frac{1}{\sqrt{N}} \sum_{j=0}^{N-1}|jjj\rangle$ is the key element of our scheme.
In recent years, experimental realizations of high-dimensional GHZ state have been reported.\upcite{GHZ1,GHZ2,GHZ3}
It is known that high-dimensional GHZ state can also be implemented by using one two-quNit CNOT gate ($\text{CNOT}|i,j\rangle=|i, i+j\; \text{mod}\; N\rangle$), one additional particle in the state $|0\rangle_a$, and generalized Bell state $|\text{Bell}\rangle=\frac{1}{\sqrt{N}} \sum_{j=0}^{N-1}|jj\rangle$. That is
\begin{eqnarray}                         \label{64}
	\begin{split}
		\text{CNOT}|\text{Bell}\rangle|0\rangle_a = |\text{GHZ}\rangle.
	\end{split}
\end{eqnarray}
Here CNOT gate with the second QuNit as the control and the ancilla quNit as the target.
Nowadays, creation of high-dimensional Bell state and implementation of CNOT gate have been studied both theoretically and experimentally.\upcite{Bell1,Bell2,CNOT1,CNOT2}
Many theoretical and experimental works have been proposed for implementing high-dimensional quantum gates, including high-dimensional CNOT gate, high-dimensional Toffoli gate, high-dimensional Pauli gate, high-dimensional Walsh-Hadamard gate, and high-dimensional Clifford + T gate.\upcite{gate1,gate2,gate3,gate4,gate5}

The projective measurement in the high-dimensional orthogonal basis is also the core component of our scheme. To clarify our description, we take $\{|\overline{\upsilon}_k\rangle\}$ described in Equation \eqref{22} as an example. The orthogonal measurement basis vector $\{|\overline{\upsilon}_0\rangle, |\overline{\upsilon}_1\rangle, |\overline{\upsilon}_2\rangle,|\overline{\upsilon}_3\rangle\}$ chosen by Charlie is given by
\begin{eqnarray}                         \label{65}
	\begin{split}
      \left(
        \begin{array}{c}
          |\overline{\upsilon}_0\rangle \\
          |\overline{\upsilon}_1\rangle \\
          |\overline{\upsilon}_2\rangle \\
          |\overline{\upsilon}_3\rangle \\
        \end{array}
      \right)
      =
		\frac{1}{2}
     \left(
    \begin{array}{cccc}
       1 & 1                             & 1                   & 1 \\
       1 & e^{\frac{\texttt{i} \pi}{2}}  & e^{\texttt{i} \pi}  & e^{\frac{\texttt{i} 3\pi}{2}}\\
       1 & e^{\texttt{i} \pi}            & e^{\texttt{i} 2\pi} & e^{\texttt{i} \pi} \\
       1 & e^{\frac{\texttt{i} 3\pi}{2}} & e^{\texttt{i} \pi}  & e^{\frac{\texttt{i} \pi}{2}} \\
    \end{array}
      \right)
      \cdot
      \left(
           \begin{array}{c}
             |0\rangle \\
             |1\rangle \\
             |2\rangle \\
             |3\rangle \\
           \end{array}
         \right).
	\end{split}
\end{eqnarray}
If we encode the qudit in the spatial mode of a single photon, i.e.,
  $|0\rangle \equiv |a_0\rangle$,
  $|1\rangle \equiv |a_1\rangle$,
  $|2\rangle \equiv |a_2\rangle$, and
  $|3\rangle \equiv |a_3\rangle$,
above transformation can be achieved by employing six variable beam splitters, see Figure \ref{Fig.2}.
The matrix forms of the variable beam splitters $T_{43}$, $T_{42}$, $T_{41}$, $T_{32}$, $T_{31}$, and $T_{21}$ can be expressed as
\begin{eqnarray}                         \label{66}
	T_{43}= \left(
	 \begin{array}{cccc}
	1 & 0                    & 0                                       & 0                                     \\
	0 & 1                    & 0                                       & 0                                     \\
	0 & 0                    & e^{\texttt{i}\phi_{43}}\sin\omega_{43}  & e^{\texttt{i}\phi_{43}}\cos\omega_{43}\\
	0 & 0                    & \cos\omega_{43}                         & -\sin\omega_{43}
   \end{array}\right),
\end{eqnarray}
\begin{eqnarray}                         \label{67}
	T_{42}= \left(
	 \begin{array}{cccc}
	1 & 0                                        & 0                    & 0                                      \\
	0 & e^{\texttt{i}\phi_{42}}\sin\omega_{42}   & 0                    & e^{\texttt{i}\phi_{42}}\cos\omega_{42} \\
	0 & 0                                        & 1                    & 0                                      \\
	0 & \cos\omega_{42}                          & 0                    & -\sin\omega_{42}
   \end{array}\right),
\end{eqnarray}
\begin{eqnarray}                         \label{68}
	T_{41}= \left(
	 \begin{array}{cccc}
  e^{\texttt{i}\phi_{41}}\sin\omega_{41}  & 0         & 0             & e^{\texttt{i}\phi_{41}}\cos\omega_{41} \\
	0                                       & 1         & 0             & 0                                      \\
	0                                       & 0         & 1             & 0                                      \\
	\cos\omega_{41}                         & 0         & 0             & -\sin\omega_{41}
   \end{array}\right),
\end{eqnarray}
\begin{eqnarray}                         \label{69}
	T_{32}= \left(
	 \begin{array}{cccc}
  1  & 0                                       & 0                                         & 0 \\
	0  & e^{\texttt{i}\phi_{32}}\sin\omega_{32}  & e^{\texttt{i}\phi_{32}}\cos\omega_{32}    & 0 \\
	0  & \cos\omega_{32}                         & -\sin\omega_{32}                          & 0 \\
	0  & 0                                       & 0                                         & 1
   \end{array}\right),
\end{eqnarray}
\begin{eqnarray}                         \label{70}
	T_{31}= \left(
	 \begin{array}{cccc}
  e^{\texttt{i}\phi_{31}}\sin\omega_{31}  & 0  & e^{\texttt{i}\phi_{31}}\cos\omega_{31}    & 0 \\
	0                                       & 1  & 0                                         & 0 \\
	\cos\omega_{31}                         & 0  & -\sin\omega_{31}                          & 0 \\
	0                                       & 0  & 0                                         & 1
   \end{array}\right),
\end{eqnarray}
\begin{eqnarray}                         \label{71}
	T_{21}= \left(
	 \begin{array}{cccc}
  e^{\texttt{i}\phi_{21}}\sin\omega_{21}  & e^{\texttt{i}\phi_{21}}\cos\omega_{21}  & 0    & 0 \\
	\cos\omega_{21}                         & -\sin\omega_{21}                        & 0    & 0 \\
	0                                       & 0                                       & 1    & 0 \\
	0                                       & 0                                       & 0    & 1
   \end{array}\right),
\end{eqnarray}
where
$\phi_{43}=\frac{\pi}{2}$,      $\omega_{43}=\frac{\pi}{4}$,
$\phi_{42}=\pi$,                $\omega_{42}=\arctan\sqrt{2}$,
$\phi_{41}=\frac{3\pi}{2}$,     $\omega_{41}=\frac{\pi}{3}$,
$\phi_{32}=\arctan(-2)$,        $\omega_{32}=\arctan\sqrt{\frac{6}{10}}$,
$\phi_{31}=\arctan(-\sqrt{2})$, $\omega_{31}=\frac{\pi}{4}$, and
$\phi_{21}=\frac{\pi}{4}$,      $\omega_{21}=\arctan(-2)$.
The subscripts $m$ and $n$ of $T_{mn}^\dag$ denote the $m^{th}$ and $n^{th}$ paths (spatial modes) of single photon, respectively.

The unitary matrix corresponding to the basis transformation in Equation \eqref{20} performed by Alice can be expressed as
\begin{eqnarray}                         \label{72}
  \begin{split}
      \left(
        \begin{array}{c}
          |\upsilon_0\rangle \\
          |\upsilon_1\rangle \\
          |\upsilon_2\rangle \\
          |\upsilon_3\rangle \\
       \end{array}
     \right)
     =
   \frac{1}{2}
    \left(
   \begin{array}{cccc}
      1 & e^{-\texttt{i}\eta_1}                              & e^{-\texttt{i}\eta_2}                    & e^{-\texttt{i}\eta_3} \\
      1 & e^{\frac{\texttt{i} \pi}{2}}e^{-\texttt{i}\eta_1}  & e^{\texttt{i} \pi}e^{-\texttt{i}\eta_2}  & e^{\frac{\texttt{i} 3\pi}{2}}e^{-\texttt{i}\eta_3}\\
      1 & e^{\texttt{i} \pi}e^{-\texttt{i}\eta_1}            & e^{\texttt{i} 2\pi}e^{-\texttt{i}\eta_2} & e^{\texttt{i} \pi}e^{-\texttt{i}\eta_3} \\
      1 & e^{\frac{\texttt{i} 3\pi}{2}}e^{-\texttt{i}\eta_1} & e^{\texttt{i} \pi}e^{-\texttt{i}\eta_2}  & e^{\frac{\texttt{i} \pi}{2}}e^{-\texttt{i}\eta_3}
   \end{array}
    \right)
    \cdot
    \left(
        \begin{array}{c}
          |0\rangle \\
          |1\rangle \\
          |2\rangle \\
          |3\rangle \\
        \end{array}
      \right).
  \end{split}
\end{eqnarray}
The unitary matrix corresponding to the basis transformation in Equation \eqref{21} performed by Bob is given by
\begin{eqnarray}                         \label{73}
  \begin{split}
      \left(
        \begin{array}{c}
          |\tilde{\upsilon}_0\rangle \\
          |\tilde{\upsilon}_1\rangle \\
          |\tilde{\upsilon}_2\rangle \\
          |\tilde{\upsilon}_3\rangle \\
       \end{array}
     \right)
     =
   \frac{1}{2}
    \left(
   \begin{array}{cccc}
      1 & e^{-\texttt{i}\tilde{\eta}_1}                              & e^{-\texttt{i}\tilde{\eta}_2}                    & e^{-\texttt{i}\tilde{\eta}_3} \\
      1 & e^{\frac{\texttt{i} \pi}{2}}e^{-\texttt{i}\tilde{\eta}_1}  & e^{\texttt{i} \pi}e^{-\texttt{i}\tilde{\eta}_2}  & e^{\frac{\texttt{i} 3\pi}{2}}e^{-\texttt{i}\tilde{\eta}_3}\\
      1 & e^{\texttt{i} \pi}e^{-\texttt{i}\tilde{\eta}_1}            & e^{\texttt{i} 2\pi}e^{-\texttt{i}\tilde{\eta}_2} & e^{\texttt{i} \pi}e^{-\texttt{i}\tilde{\eta}_3} \\
      1 & e^{\frac{\texttt{i} 3\pi}{2}}e^{-\texttt{i}\tilde{\eta}_1} & e^{\texttt{i} \pi}e^{-\texttt{i}\tilde{\eta}_2}  & e^{\frac{\texttt{i} \pi}{2}}e^{-\texttt{i}\tilde{\eta}_3}
   \end{array}
    \right)
    \cdot
    \left(
        \begin{array}{c}
          |0\rangle \\
          |1\rangle \\
          |2\rangle \\
          |3\rangle \\
        \end{array}
      \right).
  \end{split}
\end{eqnarray}

\begin{figure}[htbp]
  \centering
  \includegraphics[width=10 cm]{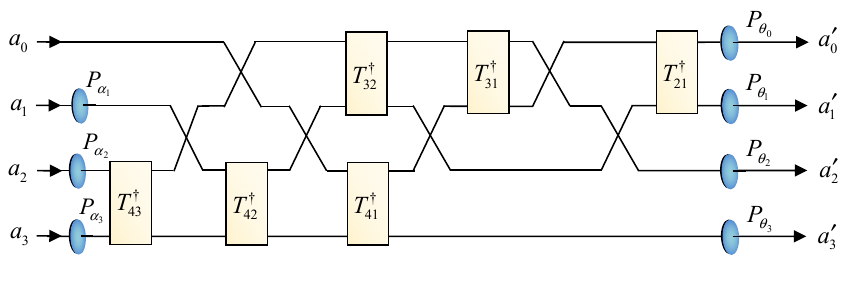}
  \caption{
  A schematic diagram of the transformation described in Equation \eqref{72} and Equation \eqref{73}. Phase shifters  $P_{\theta_0}=P_{\theta_1}=e^{\texttt{i}\arctan(-\frac{1}{3})}$, $P_{\theta_2}=e^{\frac{\texttt{i}\pi}{4}}$, and $P_{\theta_3}=e^{\frac{\texttt{i}\pi}{2}}$. $P_{\alpha_i}$ = $e^{-\texttt{i}\eta_i}$ and $P_{\alpha_i}$ = $e^{-\texttt{i}\tilde{\eta}_i}$ corresponding to Equation \eqref{72}  and Equation \eqref{73}), respectively. }
  \label{Fig.3}
\end{figure}

By adding some phase shifters to the quantum circuit of $\{|\overline{\upsilon}_k\rangle\}$, the quantum circuits for the case of $\{|\upsilon_l\rangle\}$ and $\{|\tilde{\upsilon}_n\rangle\}$ can also be designed, see Figure \ref{Fig.3}.
In Figure \ref{Fig.3}, the phase shifter $P_{\alpha_i}=e^{-\texttt{i}\eta_i}$ and $P_{\alpha_i}=e^{-\texttt{i}\tilde{\eta}_i}$ corresponding to Equation \eqref{72} and Equation \eqref{73}, respectively.

\begin{figure}[htbp]
  \centering
  \includegraphics[width=7 cm]{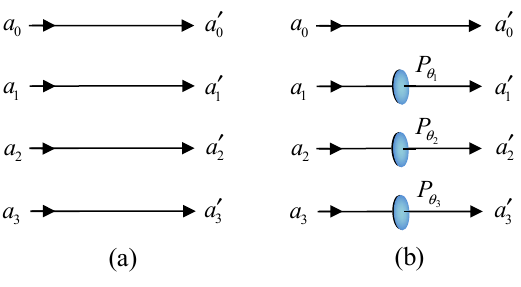}\\
  \includegraphics[width=7 cm]{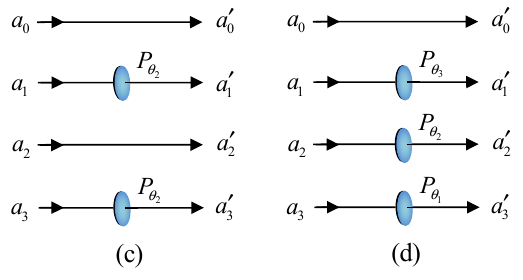}
  \caption{
  Schematic diagram of three-dimensional single-particle nontrivial operations. Phase shifters  $P_{\theta_1}=e^{\frac{\texttt{i}\pi}{2}}$, $P_{\theta_2}=e^{\texttt{i}\pi}$, $P_{\theta_3}=e^{\frac{\texttt{i}3\pi}{2}}$.}
  \label{Fig.4}
\end{figure}

It is noted that by encoding the quNit on the spatial of single photon, the single-particle nontrivial classical feed-forward operations can easy be accomplished with some phase shifters.
We take the BCRSP in four-dimensional system as an example. The quantum circuits for implementing single-qudit operations $U_0$, $U_1$, $U_2$, and $U_3$, described in Equation \eqref{28}-Equation \eqref{31}, is shown in Figure \ref{Fig.4}(a), Figure \ref{Fig.4}(b), Figure \ref{Fig.4}(c), and Figure \ref{Fig.4}(d), respectively.

In summary, we have proposed a scheme for implementing symmetric BCRSP in arbitrary high-dimensional system via two high-dimensional GHZ states.
In our scheme, Alice and Bob are both the sender and receiver, and they can deterministic remotely transmit their known single-particle high-dimensional state to each other simultaneously under the control of Charlie.
Our scheme can be completed by three steps. Firstly, Alice and Bob perform the corresponding projective measurements in the orthogonal basis on their particles $A_2$ and $B_1$, respectively.
Secondly, Alice (Bob) informs her measurement result to Bob (Alice) and Charlie by classical channels.
If Charlie wants to help them to transmit the states, he will perform single-particle measurements on his particles $C_1$ and $C_2$, respectively.
Charlie will tell Alice and Bob his measurement results by classical channels.
Finally, Alice and Bob perform the corresponding single-particle operations on their remain particles $A_1$ and $B_2$ to completely recover the original states.
Remarkably, high-dimensional CNOT gates are not required in our protocol. In previous single-qutrit (single-qudit) BCRSP,\upcite{bai2019bidirectional,shi2023controlled} six
(eight) high-dimensional CNOT gates are required via the channel with the same size.
The present scheme only considers some particular states, and more general case will be our future work.

\medskip
\section*{Acknowledgment}
This work is supported by the National Natural Science Foundation of China under Grant No. 62371038, and the Fundamental Research Funds
for the Central Universities under Grant No. FRF-TP-19-011A3.

\medskip
\section*{Conflict of Interest}
The authors declare that they have no known competing financial interests or personal relationships that could have appeared to influence the work reported in this paper.

\medskip
\section*{Data Availability Statement}
The data that support the findings of this study are available from the corresponding author upon reasonable request.

\end{document}